\pdfoutput=1
\documentclass[prl,twocolumn,
superscriptaddress,preprintnumbers,nofootinbib]{revtex4}
\usepackage{graphicx}
\usepackage{epsfig}
\usepackage{bm}
\usepackage{latexsym,amssymb,amsmath,amsfonts,amssymb,txfonts,wasysym,float}
\usepackage{scrextend}
\usepackage{mathrsfs}
\usepackage{color}

\usepackage{enumitem}

\newcommand{\postscript}[2]{\setlength{\epsfxsize}{#2\hsize}
   \centerline{\epsfbox{#1}}}

\newcommand{\mbh}{M_{\text{BH}}}

\usepackage[usenames,dvipsnames]{xcolor}
\definecolor{orange}{cmyk}{0,0.5,1,0}
\definecolor{rossoCP3}{cmyk}{0,.88,.77,.40}
\definecolor{graa}{rgb}{0.8,0.8,0.8}
\definecolor{blaa}{rgb}{0.2,0.2,0.6}

\begin{document}
\preprint{MPP-2022-125}
\preprint{LMU-ASC 29/22}

{~}

\title{\color{rossoCP3}  The Dark Universe: Primordial Black Hole $\leftrightharpoons$ Dark Graviton Gas Connection}

\author{\bf Luis A. Anchordoqui}

\affiliation{Department of Physics and Astronomy,  Lehman College, City University of
  New York, NY 10468, USA
}

\affiliation{Department of Physics,
 Graduate Center, City University
  of New York,  NY 10016, USA
}

\affiliation{Department of Astrophysics,
 American Museum of Natural History, NY
 10024, USA
}

\author{\bf Ignatios Antoniadis}
\affiliation{Laboratoire de Physique Th\'eorique et Hautes \'Energies - LPTHE
Sorbonne Universit\'e, CNRS, 4 Place Jussieu, 75005 Paris, France
}
\affiliation{Department of Mathematical Sciences, University of Liverpool
Liverpool L69 7ZL, United Kingdom}

\author{\bf Dieter\nolinebreak~L\"ust}

\affiliation{Max--Planck--Institut f\"ur Physik,  
 Werner--Heisenberg--Institut,
80805 M\"unchen, Germany
}

\affiliation{Arnold Sommerfeld Center for Theoretical Physics, 
Ludwig-Maximilians-Universit\"at M\"unchen,
80333 M\"unchen, Germany
}

\date{October 4, 2022}

\begin{abstract}
\vskip 2mm \noindent We study the possible correspondence between 5-dimensional primordial black holes and massive
5-dimensional KK gravitons as dark matter candidate 
within the recently proposed
dark dimension scenario that addresses the cosmological hierarchy
problem. 
First, we show that in the local universe a population of 5-dimensional black
holes with $M_{\rm BH}
\sim 7\times 10^{13}~{\rm g}$ would be practically indistinguishable from a KK tower of dark gravitons
with $m_{\rm DM} \sim 50~{\rm keV}$. 
Second, we connect
the mass increase  of 5-dimensional black holes and the related temperature decrease with the cooling of the tower of
massive spin-2 KK excitations of the graviton.   The dark gravitons are produced at a
mass $\sim 1 - 50~{\rm GeV}$ and the bulk of their mass shifts down to
roughly $1 - 100~{\rm keV}$ today. The cooling of the system proceeds
via  decay to lighter gravitons without losing much total mass
density, resembling the intra-tower decays that characterize the
cosmological evolution of the dynamical dark matter framework.
We associate the intra-tower
decays of the graviton gas with the black hole growth through accretion. We also discuss that the primordial black hole $\leftrightharpoons$ dark graviton gas connection
can be nicely explained by the bound state picture of black holes in terms of gravitons.
\end{abstract}
\maketitle

There is growing evidence that a vast class of quantum field theories
(QFTs), which are totally consistent as low-energy (IR) effective
theories, do not have consistent UV completions with gravity
included. Such QFTs that cannot be embedded into a UV complete quantum
gravity theory are said to reside in the Swampland, in contrast to the
effective field theories that are low energy limits of string theory
and inhabit the Landscape~\cite{Vafa:2005ui}. This sorting of QFTs by
their consistency with gravity has become an unexpectedly powerful
theoretical tool, offering potential solutions to the problems of
fine-tuning~\cite{Palti:2019pca,vanBeest:2021lhn}. 

To be specific, the distance
conjecture points to the emergence of infinite towers of states that become
light and imply a breakdown of the QFT in the infinite distance limits
in moduli space~\cite{Ooguri:2006in}.
Stated in the form of the AdS
distance conjecture, it  suggests 
that a small cosmological constant corresponds to an
an infinite field distance limit in field space and
that
there should be an infinite tower of states, whose mass is related to
the magnitude of the cosmological
constant~\cite{Lust:2019zwm}. 
The spectacular phenomenological consequences of this assertion 
in connection with the  smallness of the
cosmological constant in Planck units ($\Lambda \sim 10^{-122} M_{\rm
  Pl}^4$) has then recently led to the proposal of the dark universe scenario~\cite{Montero:2022prj}. It is precisely in this asymptotic limit
where quantum gravity (QG) effects become important at scales much
below $M_{\rm Pl}$ and hence QG can have an effect on the IR. Indeed, physics would become strongly coupled to
gravity at the species scale $M_{\rm UV}$, which corresponds to
 the Planck scale of the higher dimensional theory~\cite{Dvali:2007hz,Dvali:2007wp}.

Specifically for the dark universe,  requiring the experimental limits (on deviations from Newton's gravitational inverse-square
 law~\cite{Lee:2020zjt} and neutron star heating~\cite{Hannestad:2003yd}) to
 be consistent with the theoretical bound from the swampland
 conjectures leads to the prediction of a single extra
mesoscopic dimension (a.k.a. the dark dimension) of length $R_\perp
\sim \lambda \Lambda^{-1/4}$, with $10^{-1} \alt \lambda \alt
10^{-4}$~\cite{Montero:2022prj}. The dark dimension
opens up at the characteristic mass scale of the tower, where physics must be
described by an QFT in higher dimensions up to the species
scale, $M_{\rm UV} \sim \lambda^{-1/3} \Lambda^{1/12} M_{\rm
  Pl}^{2/3}$.

This seemingly simple model is behind a
very rich
phenomenology~\cite{Montero:2022prj,Anchordoqui:2022ejw,Anchordoqui:2022txe,Blumenhagen:2022zzw,Gonzalo:2022jac}. For
example, two dark matter (DM) candidates have been proposed: {\it (i)}~primordial
black holes (PBHs) that were born very early in the life of the
universe~\cite{Anchordoqui:2022txe}, and {\it (ii)} the
massive spin-2 KK excitations of the graviton in the dark dimension
(dubbed the ``dark gravitons'')~\cite{Gonzalo:2022jac}. In this Letter
we investigate whether there is a possible connection between these DM
candidates. Our investigation is deep-rooted on the fact that, for many purposes, a black hole can be replaced by a bound state of gravitons~\cite{Dvali:2011aa}.\footnote{This possible
connection was already mentioned in ~\cite{Gonzalo:2022jac}.}

For both  scenarios, 
we consider the canonical picture featuring D-brane-localized fields of
the Standard Model (SM) with only gravity propagating in the 5-dimensional bulk~\cite{Antoniadis:1998ig}. 
Let us first  recall some of the basic features of the 5-dimensional PBHs scenario for dark matter, as introduced in~\cite{Anchordoqui:2022txe}.
The radius of a $(4+n)$-dimensional Schwarzschild black hole is given as
\begin{equation}
r_s(\mbh) =
\frac{1}{M_{{\rm Pl},n}}
\left[ \frac{\mbh}{M_{{\rm Pl},n} } \,\, \frac{2^n \pi^{(n-3)/2}\Gamma({n+3\over 2})}{n+2}
\right]^{1/(1+n)}\,,
\label{Sch}
\end{equation}
with the $(4+n)$ Planck mass
\begin{equation}
M_{{\rm Pl},n} =     \left(\frac{R_\perp^{-n} \ M_{\rm Pl}^2}{8 \pi}\right)^{1/(n+2)}   \,, 
\end{equation}
where $R_\perp$ is the radius of the extra dimensional compact space, and
where 
$\Gamma(x)$ is the
Gamma function~\cite{Myers:1986un}.
The transition between a 4-dimensional black hole and a $(4+n)$-dimensional occurs when $r_s(\mbh)<R_\perp$,
and this transition occurs independent from $n$ at  black hole masses $\mbh< R_\perp \ M_{\rm Pl}^2$.
Note that at the transition point the four-dimensional Schwarzschild radius agrees with the higher dimensional one.
As observed in~\cite{Anchordoqui:2022txe}, for $R_\perp\simeq
\lambda \, \Lambda^{-1/4}$ and $n=1$ as suggested in the dark universe~\cite{Montero:2022prj},
there is a remarkable coincidence, namely the 4D-5D transition occurs at a black hole mass of about 
$\mbh\sim 10^{20}~{\rm g}$, precisely the value, below which primordial black holes are viable all-dark-matter candidates.

PBHs radiate all particle species lighter than or comparable to
their temperature $T_{\rm BH}$. For a higher-dimensional black the Hawking temperature is given as~\cite{Anchordoqui:2001cg} 
\begin{equation}
T_{\rm BH} = \frac{n +1}{4\,\pi\,r_s} \, ,
\label{TBH}
\end{equation}
and the associated entropy takes the form
\begin{equation}
  S = \frac{4 \, \pi\,\mbh\,r_s}{n+2} \,.
\end{equation}

\begin{figure}[tpb]
\postscript{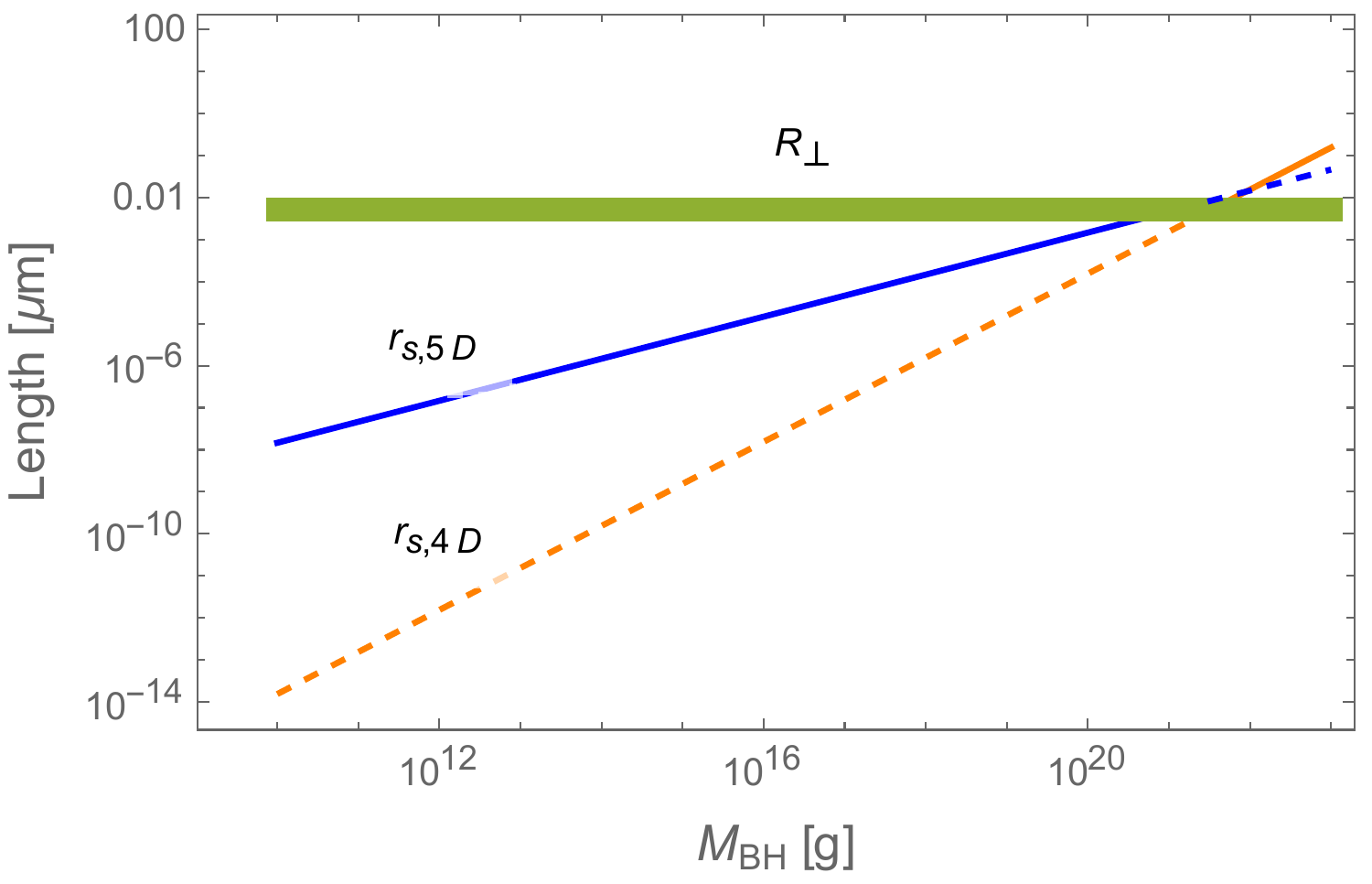}{0.99}
\caption{Scaling of the Schwarzschild radius with $M_{\rm BH}$ for 4-
  and 5-dimensional  black holes. The thick horizontal line indicates the size of the
  compactification radius. \label{fig:1}}
\end{figure}

\begin{figure}[tpb]
\postscript{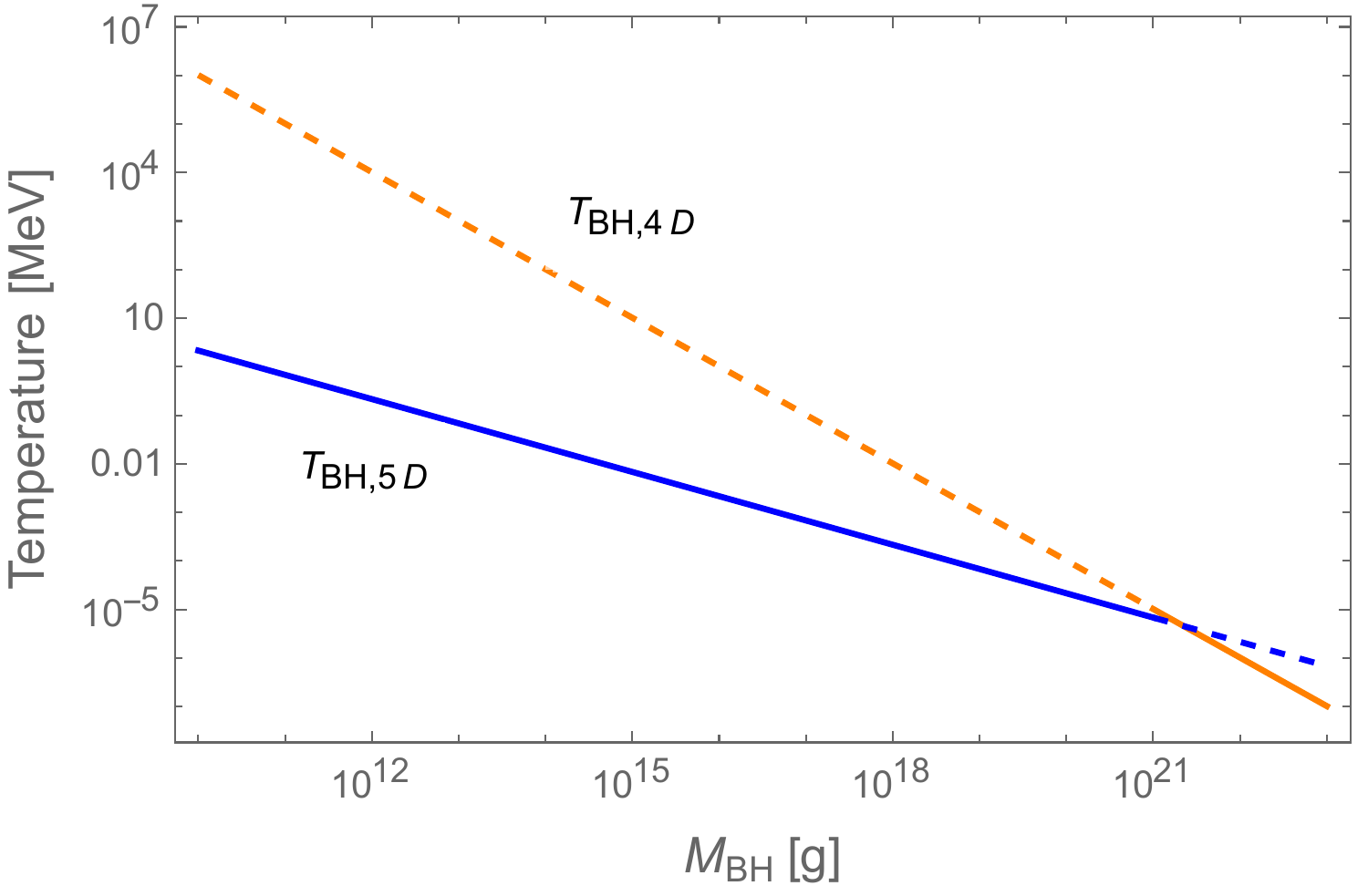}{0.99}
\caption{Scaling of $T_{\rm BH}$ with $M_{\rm BH}$ for 4- and
  5-dimensional 
  black holes. \label{fig:2}}
\end{figure}

In Figs.~\ref{fig:1}
and \ref{fig:2} we show the (logarithmic) scaling of the Schwarzschild radius and
$T_{\rm BH}$ with $M_{\rm BH}$ for 4- and 5-dimensional black
holes. For $M_{\rm BH} \sim 10^{20}~{\rm g}$, the horizon size is
equal up to some numerical constants to the compactification radius. 
As one can see, for a given black hole mass, the five-dimensional horizon is larger than for a four-dimensional black hole, whereas the Hawking temperature is smaller in five dimensions than
in four dimensions. As a result the five-dimensional black holes have a longer life time compared to four-dimensions, which makes them  viable all-dark-matter candidates.

Let us briefly also recall the main features of the dark gravitons as dark matter candidates, where we refer to~\cite{Gonzalo:2022jac}
for more details.
The dark gravitons are produced
thermally by the hot brane at $m_{\rm DM} \sim 4~{\rm GeV}$. While 
the total mass of the dark KK gravitons remains approximately constant
over time, its mass distribution quickly shifts to lower values by the
large number of available KK modes of lighter mass, thereby preventing
DM from substantially decaying back to SM fields. The dark graviton
model then provides a particular realization of the dynamical dark
matter  (DDM)
framework~\cite{Dienes:2011ja}, with dominance of intra-tower decays
and a small decay rate ($\propto m_{\rm DM}^3$) to the SM
fields. Consistency with experimental data requires that the DM mass
from the time of matter-radiation (MR) equality till today decreases
from $m_{\rm DM} ({\rm MR}) \sim
1~{\rm MeV}$ to $m_{\rm DM} ({\rm today}) \sim 50~{\rm keV}$~\cite{Gonzalo:2022jac}. 

Before proceeding, we pause to review experimental
  constraints on these dark matter candidates. PBH dark matter has
  been challenged by many observational probes that set bounds on the
  relative PBH abundance across a broad range of viable PBH
  masses~\cite{Villanueva-Domingo:2021spv}. Among these bounds, those coming from microlensing
  gravitational surveys yield the most restrictive constraints for
  $M_{\rm BH} \agt 10^{22}~{\rm g}$~\cite{Croon:2020ouk}. An
  astonishing coincidence is that the size of the dark dimension
  $R_\perp \sim$ wavelength of visible light. This means that the
  Schwarzschild radius of 5D black holes is well below the wavelength
  of light. For point-like lenses, this is the critical length where
  geometric optics breaks down and the effects of wave optics suppress
  the magnification, obstructing the sensitivity to 5D PBH
  microlensing signals. In principle, femtolensing of cosmological gamma-ray bursts (GRBs) could be used to search for dark matter
  objects in the mass range
  $10^{17} \alt M_{\rm BH}/{\rm g} \alt
  10^{20}$~\cite{Gould:1992}. The lack of
  femtolensing detection by the {\it Fermi} Gamma-ray Burst Monitor was used
  to constrain PBHs with mass in the range $10^{17.7} < M_{\rm
    BH}/{\rm g} \alt 10^{20}$ to
  contribute no more than 10\% to the total dark matter
  abundance~\cite{Barnacka:2012bm}. However, the validity of this GRB constraint has been called
  into question,
  because it is based on the assumption that the gamma-ray source is
  point-like in the lens plane~\cite{Katz:2018zrn}. The non-pointlike nature of GRBs imply that most of them are too big when projected onto the
  lens plane to yield meaningful femtolensing limits. Indeed, only a
  small GRB population with very fast variability might be suitable for PBH
  searches. When this systematic effect is taken into account to
  decontaminate the {\it Fermi} 
  sample, the GRB constraint on PBHs becomes obsolete. A sample of 100
  GRBs with transverse size $10^9~{\rm cm}$ would be needed to probe
  the 5D PBHs~\cite{Katz:2018zrn}. Should a positive signal be detected, it can be
  used as a discriminator between the
  PBH and dark graviton hypotheses.

A combination of temperature and polarization data from {\it Planck}
provides strong constraints on the abundance of dark gravitons and
PBHs. On the one hand, the model-independent bound on decaying dark
matter derived in~\cite{Slatyer:2016qyl} constrains the KK mass $\alt 1~{\rm MeV}$ at the epoch
of matter-radiation equality~\cite{Gonzalo:2022jac}. On the other
hand, {\it Planck} data combined with observations of the MeV extragalactic gamma-ray background
 constrain the abundance of PBHs for $M_{\rm BH} \alt 10^{12}~{\rm g}$.\footnote{This mass
   range has been estimated by a rudimentary rescaling of the bound of derived
   in~\cite{Clark:2016nst} for 4D PBHs.}

Now let us explain the possible microscopic reason for the correspondence between the black holes and the dark gravitons as dark matter candidates.
It is based on the description~\cite{Dvali:2011aa}, where a black hole can be viewed as being composed, i.e. is a bound state, of $N$ gravitons.
The number $N$ then agrees with the entropy of the black hole: $N\simeq S$. So far this picture was used for massless gravitons.
But we now apply it to a black hole of mass $\mbh$, which is composed of $N$ massive gravitons, each  of mass $m_{\rm DM}$.
So we can immediately determine the number $N$, i.e. the number of massive graviton constituents, as 
\begin{equation}
N = {\mbh\over m_{\rm DM}}\, .
\end{equation}
As said, $N$ agrees with the entropy, and we can also express the entropy as the ratio of black mass and temperature, i.e.
\begin{equation} 
N\simeq S\simeq {\mbh\over T_{\rm BH}}\, .
\end{equation}

\vskip0.3cm
\noindent
{\sl From that we first deduce that we should associate the graviton constituent mass $m_{\rm DM}$ with the black hole Hawking temperature $T_{\rm BH}$:}
\begin{equation}
m_{\rm DM}\simeq  T_{\rm BH}\, .
\end{equation}
\vskip0.3cm

As said, the black hole contains $\mbh/m_{\rm DM}$ dark gravitons and the total dark matter mass density  is the same in both pictures.
Furthermore the decay of the higher mass  KK dark gravitons to lower mass gravitons corresponds to the accretion of the black hole, i.e. to
lowering its temperature by increasing its mass. The black hole entropy also increases by the accretion process, so the number $N$ of dark gravitons also increases accordingly
and the overall mass stays the same.
This brings us  to a second, new interesting proposal for the black hole
$\leftrightharpoons$ graviton gas correspondence: 

\vskip0.3cm
\noindent
{\it The cooling of the
graviton gas can be visualized as the black hole growth by accreting
mass from the surrounding plasma}. 
\vskip0.3cm

This is an interesting association
that we now turn to study in detail for the dark universe. Here the Hawking temperature is related   to the black
hole mass $M_{\rm BH}$ by~\cite{Anchordoqui:2022txe}
\begin{eqnarray}
  T_{\rm BH} & = & \sqrt{\frac{3}{64}} \ \frac{1}{\pi} \ \frac{M_{\rm Pl} \
    \Lambda^{1/8}}{\lambda^{1/2} \ M_{\rm BH}^{1/2}} \nonumber \\ & \approx &
    \left(\frac{M_{\rm BH}}{5 \times 10^{10}~{\rm g}}\right)^{-1/2}~{\rm MeV} \, ,
\end{eqnarray}
where we have taken $\lambda = 10^{-4}$. This radiation causes PBHs to lose mass at a rate
\begin{eqnarray}
 \left. \frac{dM_{\rm BH}}{dt}\right|_{\rm evap} & = & -\frac{\Lambda^{1/4} \ M_{\rm
                               Pl}^2}{640 \
      \pi \ \lambda \ M_{\rm BH}} \ \sum_{i} c_i(T_{\rm BH}) \ \tilde f \ \Gamma_s
    \nonumber \\
& = & - 2.7 \times 10^{27}~{\rm GeV}^3 \frac{1}{M_{\rm BH}} \sum_{i}
      c_i (T_{\rm BH})  \tilde f
  \Gamma_s  \,,     
\label{rate}
\end{eqnarray}
where $c_i(T_{\rm BH})$ counts the number of internal degrees of freedom of particle 
species $i$ of mass $m_i$ satisfying $m_i \ll T_{\rm BH}$,  $\tilde f = 1$  $(\tilde f = 7/8)$ for bosons
(fermions), and where $\Gamma_{s=1/2} \approx 2/3$ and $\Gamma_{s=1} \approx
1/4$ are the (spin-weighted) dimensionless greybody factors normalized to the black
hole surface area~\cite{Anchordoqui:2022txe}. We are interested here in
$T_{\rm BH} \alt 1 {\rm MeV}$ and so $c_i(T_{\rm BH})$ receives a contribution of 6 from neutrinos, 4 from electrons, 2 from photons.
In the spirit of~\cite{Emparan:2000rs}, we neglect graviton emission because the KK modes are excitations in
the full transverse space, and so their overlap with the small
(higher-dimensional) black holes is suppressed by the geometric factor
$(r_s/R_\perp)^2$ relative to the brane fields. Thus, the geometric suppression precisely compensates for the enormous number of modes, and the total contribution of all KK modes is only the same order as that from a single brane field.

Integrating (\ref{rate}) we find that a black hole with an initial mass of
$M_{\rm BH} \sim  2 \times 10^{11}~{\rm g}$ will evaporate in a time equal to the age of
the universe (which is taken to be $13.787 \pm 0.020~{\rm  Gyr}$~\cite{Planck:2018vyg}). The characteristic energy of
the emitted particles is 500~keV. A black hole of $M_{\rm BH} \sim 5
\times 10^{12}~{\rm g}$
radiates particles with an average energy $\sim
100~{\rm keV}$ and has a lifetime $\sim 7.6 \times 10^{10}~{\rm yr}$, whereas
for $M_{\rm BH} \sim 7 \times 10^{13}~{\rm g}$, the average particle
energy $\sim 25~{\rm keV}$ and the black hole lifetime $\sim 1.5 \times
10^{15}~{\rm yr}$. On the other hand, PBHs of $M_{\rm BH} \sim 5 \times 10^{10}~{\rm g}$  have a lifetime of 0.5~Gyr
and radiate particles with average energy $\sim 1~{\rm MeV}$. 

In order to compare with the behaviour of the graviton dark matter candidates, we now turn to ascertain whether 5-dimensional PBHs of
$T_{\rm BH}({\rm MR}) \sim 1~{\rm MeV}$ at the MR equality could cool down to
$T_{\rm BH} ({\rm today})\sim 25~{\rm keV}$ and  survive until today. The net change of the black hole mass is given by a balance equation
\begin{equation}
\frac{dM_{\rm BH}}{dt} = \left. \frac{dM_{\rm BH}}{dt}\right|_{\rm
  accr} + \left. \frac{dM_{\rm BH}}{dt}\right|_{\rm evap} \,\, ,
\label{balance}
\end{equation}
where 
\begin{equation}
\left. \frac{dM_{\rm BH}}{dt}\right|_{\rm accr} \approx \frac{64
  \pi}{3}  \ \frac{M_{\rm BH} \ \lambda}{\Lambda^{1/4} \ M_{\rm Pl}^2}
\ \varepsilon,
\, ,
\label{accr}
  \end{equation}
and where $\varepsilon$ is the energy density of the plasma in the
vicinity of the event horizon~\cite{Anchordoqui:2022txe}. Substituting (\ref{rate}) and (\ref{accr}) into (\ref{balance}), it follows that for  accretion to
dominate the black hole evaporation,
\begin{eqnarray}
  \varepsilon & > & \frac{1}{40960 \ \pi^2} \ \frac{\Lambda^{1/2} \ M_{\rm
      Pl}^4}{\lambda^2 \ M_{\rm BH}^2} \sum_i c_i \ \tilde f \ \Gamma_s
  \nonumber \\ 
& > & \frac{2 \times 10^{56}}{M_{\rm BH}^2} \ \sum_i c_i \ \tilde f \
  \Gamma_s~{\rm GeV}^6 \, .
\end{eqnarray}
For $M_{\rm BH} \sim 5 \times
10^{10}~{\rm g}$, we find $\epsilon > 2 \times 10^{-10}~{\rm
  GeV/fm}^3$. Amusingly, for the temperature range of interest
($T_{\rm BH} \alt 1~{\rm MeV}$) the required $\varepsilon$ for black hole stability is well below the energy density of partons formed
in lead-lead collisions at LHC: $\varepsilon_{\rm LHC} \sim 500~{\rm
  GeV/fm^3}$~\cite{Chamblin:2003wg}. 

\begin{table}[t]
\caption{History of temperature changes in the early universe~\cite{Workman:2022ynf}. \label{tabla1}}
  \begin{tabular}{|c|c|c|}
\hline \hline
    Epoch & Temperature & Age \\
\hline                          
    Electroweak Phase Transition & $T \sim 100~{\rm GeV}$ & $20~{\rm ps}$ \\
    QCD Phase Transition & $T \sim 150~{\rm MeV}$ & $20~\mu{\rm s}$ \\
    Neutrino Decoupling & $T \sim 1~{\rm MeV}$ &  $ 1~{\rm s}$\\
Electron-Positron Annihilation & $T<m_e < 0.5~{\rm MeV}$ & $
                                                          10~{\rm s}$
    \\
    Big Bang Nucleosynthesis & $50 \alt T/{\rm keV} \alt 100 $ & $10~{\rm min}$
    \\
    Matter-Radiation Equality & $T\sim 0.8~{\rm eV} \sim 9000~{\rm K}$
                        & $6 \times 10^4~{\rm yr}$
    \\
    Photon Decoupling & $T \sim 0.3~{\rm eV} \sim 3000~{\rm K}$ &
    $ 3.8 \times 10^5~{\rm yr}$ \\
\hline \hline
  \end{tabular}
  \end{table}

Next, in line with our stated plan, we  get an estimate of the average
energy density
in the primordial plasma by looking at the radiation energy density,
which is given by the Stefan-Boltzmann law
\begin{equation}
 \langle \varepsilon_{\rm rad} \rangle = \frac{\pi^2}{30} \ g_\varepsilon(T) \ T^4 \,,
\end{equation}  
where $g_\varepsilon(T)$ counts the effective number of relativistic
helicity degrees of freedom at a given photon temperature $T$; as in
(\ref{rate}) fermionic degrees of freedom are suppressed by a factor
7/8 with respect to bosonic degrees of freedom. Using the
characteristic temperatures for the different epochs of the universe
given in Table~\ref{tabla1} we find that at MR equality
$\langle \varepsilon_{\rm rad} \rangle \sim 2 \times 10^{-34}~{\rm GeV/fm}^3$, whereas at neutrino decoupling
$\langle \varepsilon_{\rm rad} \rangle \sim 5 \times 10^{-10}~{\rm
  GeV/fm}^3$. We conclude that the accretion of the black hole and the cooling down process can generically occur at $T\sim 1~ {\rm MeV}$, i.e. at the epoch,
where the neutrinos are decoupling. However one can imagine to push down this temperature to the epoch of matter-radiation equality. Since the average density is already much lower at this epoch,
 it may be possible to cool down a black hole from $T_{\rm BH} ({\rm MR}) \sim 1~{\rm MeV}$
 to $T_{\rm BH} ({\rm today})\sim 25~{\rm keV}$, only if the primordial
plasma at matter-radiation equality has fluctuations at the
level of $\sim 1$ in $10^{24}$. Whether this is possible, is an open and model dependent question.

In summary, we have discussed the possible correspondence 
between dark gravitons and primordial black holes as dark matter candidates in the dark universe.
Associating the black hole Hawking temperature with the mass of the dark graviton, 
we have shown that today a population of 5-dimensional PBHs with $M_{\rm BH}
\sim 7\times 10^{13}~{\rm g}$ would be practically indistinguishable from a KK tower of dark gravitons
with $m_{\rm DM} \sim 50~{\rm keV}$. 
We have shown that up to some numerical factors  the radiation rate of
5-dimensional black holes is comparable to the graviton emission into
brane-localized SM fields. Although this is not the dominant channel describing the cosmological evolution of the dark universe,
it is the one that produces visible signals for observers living on
the brane. In other words, from the point of view of the 4-dimensional
observers it will be challenging to distinguish among the two DM
candidates. 

In addition we have investigated the possible correspondence between the
mass change rate of 5-dimensional black holes and the cooling of the tower of massive spin-2 KK excitations of the
graviton in the dark dimension. The dominant decay channel of the dark graviton gas is into lighter gravitons. This has a correspondence with the black
hole accretion process.
The dark gravitons are produced at a
mass $\sim 1 - 50~{\rm GeV}$ and the bulk of their mass shifts down to
roughly $1 - 100~{\rm keV}$ today. The cooling of the system proceeds
via  decay to lighter gravitons without losing much total mass
density, resembling the intra-tower decays characterizing the
cosmological evolution of the DDM framework. 
We interpreted the intra-tower
decays of the graviton gas with the black hole growth via accretion
of the plasma in the vicinity of its horizon. In principle, with the right amount of energy
density on the plasma surrounding the black hole horizon one could
make a one-to-one correspondence. However, it happens to be that the
cosmic expansion of our universe does not have {\it on average} enough energy density
for black holes with $T_{\rm BH} \sim 1~{\rm MeV}$ to overtake the
radiation process via accretion at the matter radiation
equality.  It is around the epoch of neutrino decoupling where the
average energy density of the plasma may allow black holes of $T_{\rm BH} \sim 1~{\rm MeV}$ to survive until today via accretion.

\section*{Acknowledgments}

We thank Cumrun Vafa for useful comments on the manuscript. The work of L.A.A. is supported by the U.S. National Science
Foundation (NSF Grant PHY-2112527). The work of D.L. is supported by the Origins
Excellence Cluster and by the German-Israel-Project (DIP) on Holography and the Swampland.

\end{document}